\documentclass[journal]{IEEEtran}
\ifCLASSINFOpdf
\else
\fi

\usepackage{graphicx}
\usepackage{dcolumn}
\usepackage{bm}
\usepackage[caption=false]{subfig}
\begin{document}
	%
	\title{A Thermodynamic Perspective of Negative-capacitance Field-effect-transistors}
	%
	%
	%
	
	\author{Sou-Chi Chang,~Uygar E. Avci,~Dmitri E. Nikonov~\IEEEmembership{Senior Member,~IEEE,}~and~Ian A. Young~\IEEEmembership{Fellow,~IEEE}
		\thanks{S. -C. Chang, U. E. Avci, D. E. Nikonov, and I. A. Young are with Components Research, Intel Corporation, Hillsboro, OR 97124, USA.}}
	
	%
	%

	\markboth{JxCDC template}%
	{Shell \MakeLowercase{\textit{et al.}}: Bare Demo of IEEEtran.cls for IEEE Journals}
	%



	\maketitle
	
	\begin{abstract}
		Physical phenomena underlying operation of ferroelectric field-effect transistors (FeFETs) is treated within a unified simulation framework. The framework incorporates the Landau mean-field treatment of free energy of a ferroelectric and the polarization dynamics according to Landau-Khalatnikov (LK) equation. These equations are self-consistently solved with the one-dimensional metal-oxide-semiconductor (MOS) structure electrostatics and the drift-diffusion solution for the current in the semiconductor channel. Numerical simulations demonstrate, depending on the ferroelectric (FE) thickness, both regimes of hysteresis switching (relevant for a non-volatile memory) and of higher on-currents and steeper subthreshold slope (SS) with a negligible hysteresis (relevant for logic) via the negative capacitance effect.
	\end{abstract}
	

	%
	\IEEEpeerreviewmaketitle

	\section{Introduction}
	Over the past four decades, the computing power of microprocessors has been exponentially increasing thanks to the relentless pursuit of Moore's law, stating that the number of transistors in an integrated circuit has doubled approximately every two years \cite{658762}. However, as the complementary metal-oxide-semiconductor (CMOS) technology is scaled to single nanometer sizes, the static power component becomes increasingly dominant share of total energy dissipation due to the reduction of the on-off current ratio ($\frac{I_{on}}{I_{off}}$) in CMOS transistors \cite{1250885}. It has been well known that the high on-off current ratio can be achieved by minimizing the subthreshold swing of a transistor, which is defined as
	\begin{eqnarray}
	SS = \left(\frac{\partial \psi_{s}}{\partial V_{g}}\frac{\partial \log_{10}I_{ds}}{\partial \psi_{s}}\right)^{-1}, 
	\end{eqnarray}
	where $\psi_{s}$ and $V_{g}$ are the surface potential of the transistor's channel and the gate voltage, respectively, and $I_{ds}$ is the source-to-drain current. In the expression of SS, 
	\begin{eqnarray}
	E_{s} &=& \frac{\partial \psi_{s}}{\partial V_{g}}, \\
	C_{s} &=& \frac{\partial \log_{10}I_{ds}}{\partial \psi_{s}}
	\end{eqnarray} 
	are the factors related to the electrostatic control in a MOS capacitor and the amount of current that can be provided by the channel's band structure, respectively. In CMOS transistors, the upper limit of $E_{s}$ is $1$ because the gate voltage has to be dropped in each layer of the stack - the high-k dielectric, SiO$_{x}$ under-layer, and semiconductor channel. The upper limit of $C_{s}$ is typically $ 2.3\frac{k_{B}T}{e}$ since the transport mechanism in the subthreshold region is dominated by thermionic emission of carriers from the source terminal. As a consequence, the minimum of $SS$ that can be achieved in a CMOS transistor is $\sim 60$mV/dec at room temperature. 
	
	To drive $SS$ below $60$mV/dec at room temperature, tunnel field-effect transistors (TFETs) have been proposed to improve $C_{s}$ by changing the channel conduction mechanism from diffusive transport to quantum-mechanical tunneling \cite{1474084,Theis2010,Ionescu2011}. Nevertheless, there are still several issues associated with TFETs such as low on currents, relatively complicated process flow, and needs for circuit schematics modifications. On the other hand, recently achieving $E_{s} > 1$ has been extensively explored by several research groups by way of the negative capacitance (NC) effect of the ferroelectric (FE). In these devices, known as NCFETs \cite{doi:10.1021/nl071804g}, the only required modification is replacing high-k dielectric (DE) with the FE oxide. The main idea of NCFETs can be understood in a fairly simple way as follows. For a gate with the ferroelectric layer, the channel structure factor can be expressed as 
	\begin{eqnarray}
	E_{s} = \left(1+\frac{C_{semi}}{C_{ox}}\right)^{-1}, 
	\end{eqnarray}
	where $C_{semi}$ and $C_{ox}$ are the capacitors associated with semiconductor and gate oxide, respectively. Then it is possible to have $E_{s}>1$ if $C_{ox}$ becomes negative.
	
	Over the past few years, NC in the FE has been under extensive discussion \cite{doi:10.1021/nl071804g,Khan2015,Zubko2016,Krowne2011,Catalan2015}. Some groups claimed that NC originates from the down-pointing curvature of energy barrier between two stable polarization states in the thermodynamic free energy profile and can be directly measured during polarization reversals in a FE capacitor \cite{doi:10.1021/nl071804g,Khan2015}. Others argued that NC observed during polarization reversals is a pure electrostatic effect, rather than thermodynamic one \cite{Catalan2015}. As will be explained in details below, observing NC directly in a single FE capacitor is actually prohibited by laws of thermodynamics; however, NC effects in the FE do result from the negative curvature of energy barrier and can still be deduced from the enhancement of overall capacitance as a FE capacitor is in series with other capacitors.
	
	\begin{figure*}
		\begin{center}
		\subfloat[]{%
			\includegraphics[width=.12\linewidth]{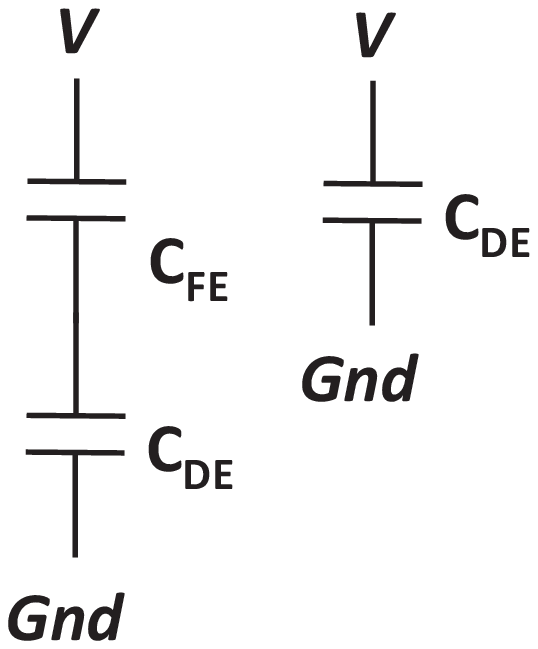}%
		}
		\subfloat[]{%
			\includegraphics[width=.35\linewidth]{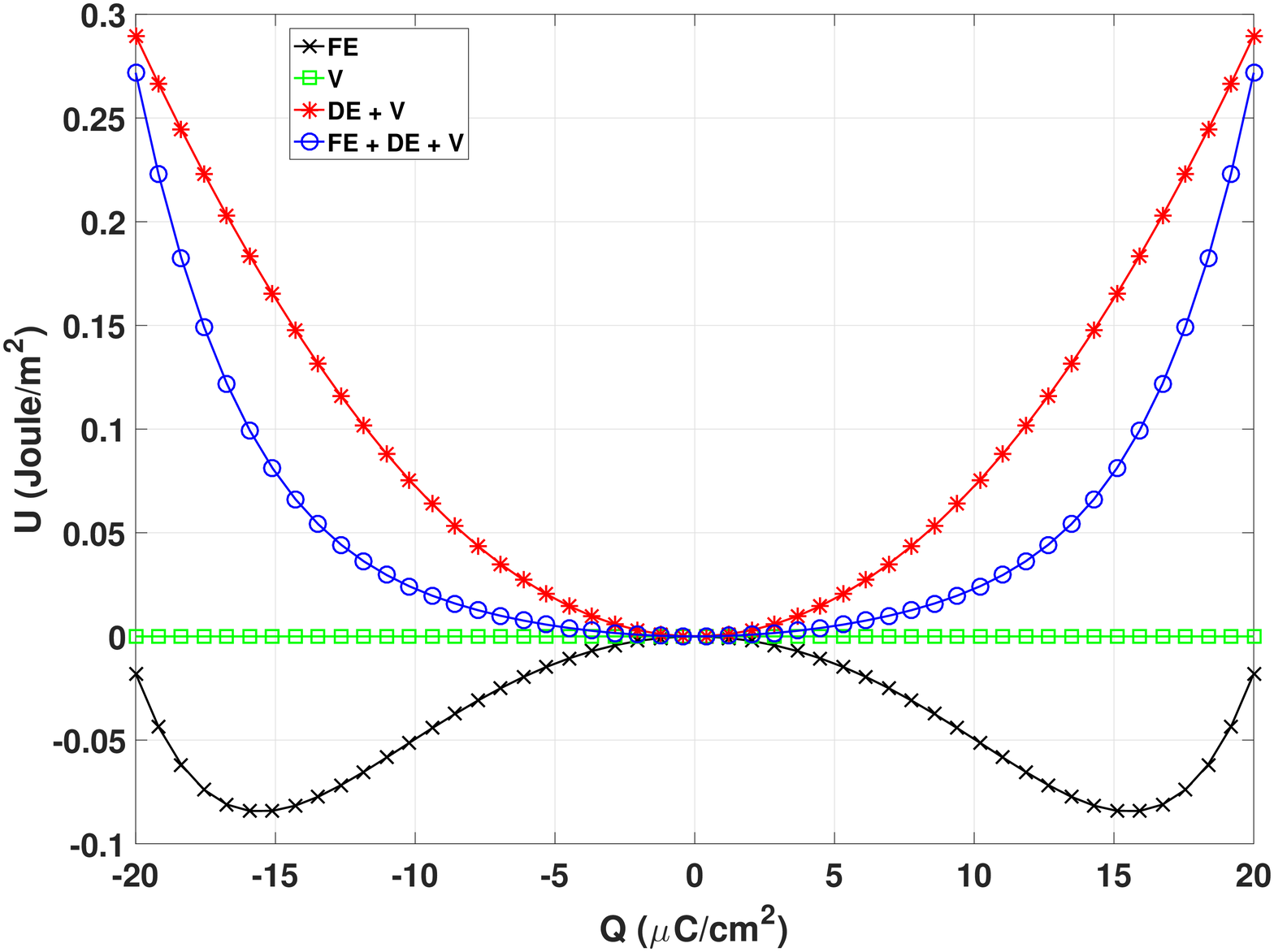}%
		}\\
		\subfloat[]{%
			\includegraphics[width=.35\linewidth]{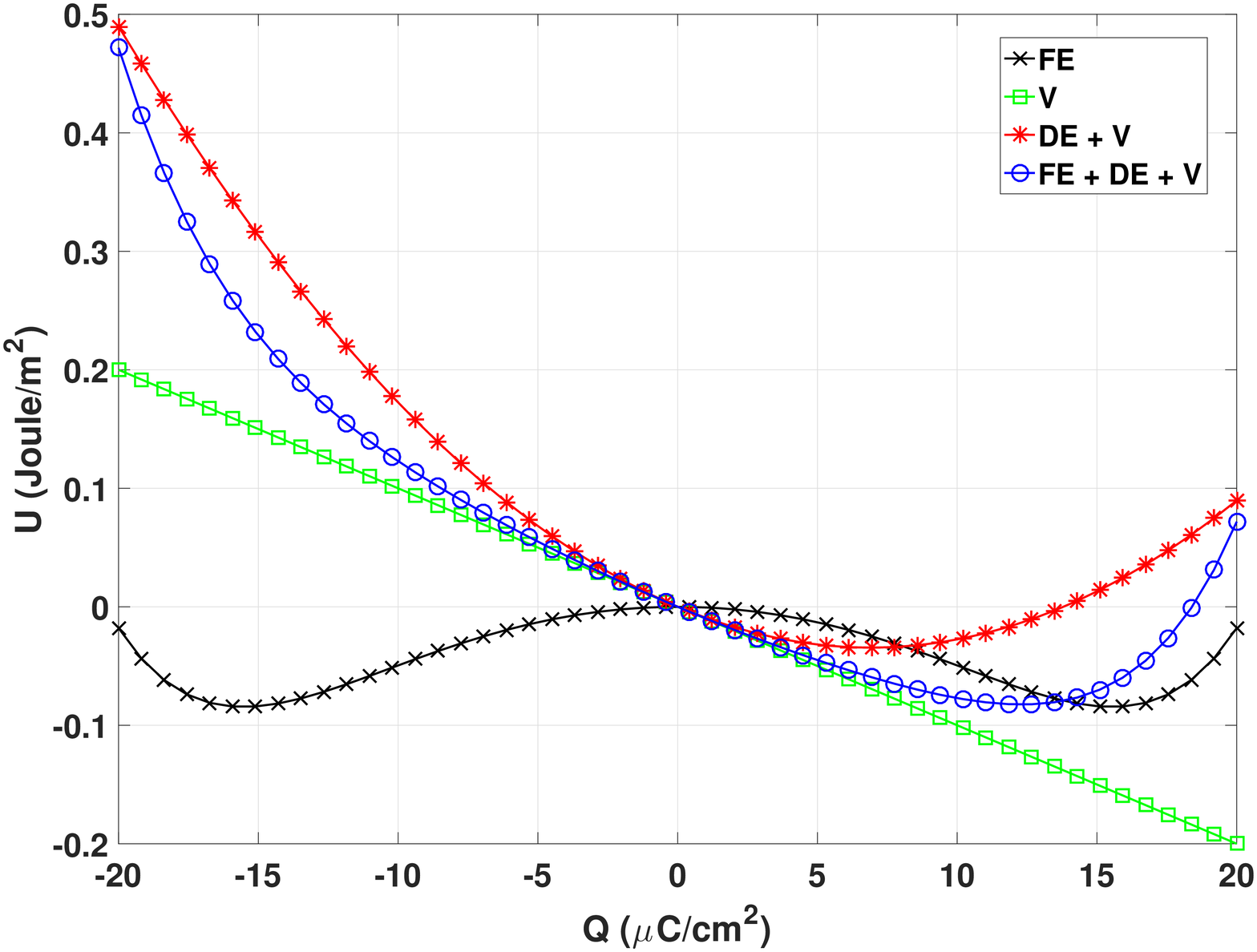}%
		}
		\subfloat[]{%
			\includegraphics[width=.35\linewidth]{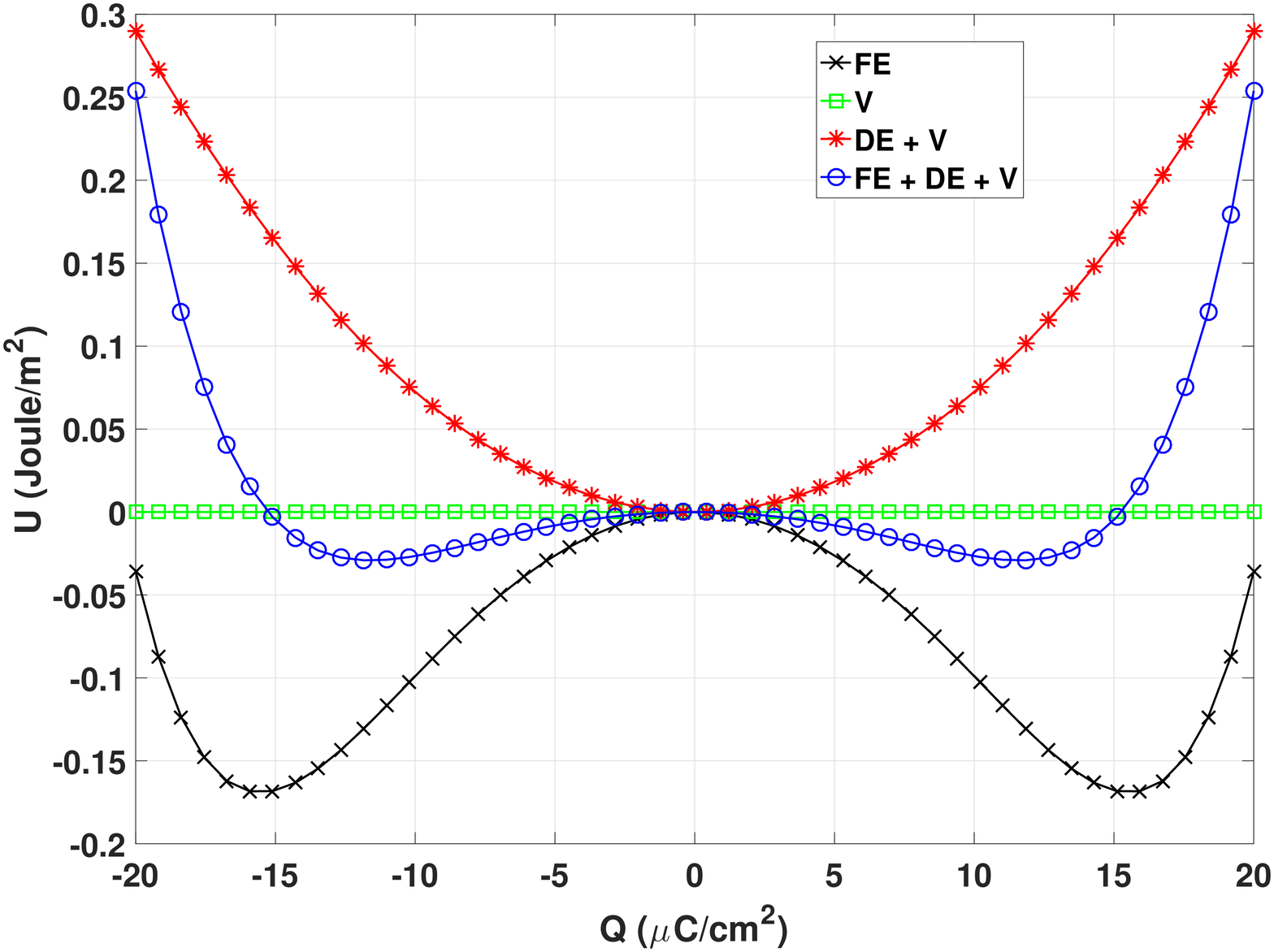}%
		}
		\caption{(a) Schematics of DE and FE capacitors in series as well as a single DE capacitor. Free energy profiles of normal DE (red curve) and DE-FE (blue curve) capacitors with (b) $t_{FE}=5$nm, $t_{DE}=0.5$nm, $V=0$V, (c) $t_{FE}=5$nm, $t_{DE}=0.5$nm, $V=1$V, and (d) $t_{FE}=10$nm, $t_{DE}=0.5$nm, $V=0$V. The green and black curves designate contributions to the free energy by the applied voltage and the FE capacitor, respectively.}
		\label{fig1}
		\end{center}
	\end{figure*}
	
	In addition to the promise of the NC effect to improve $\frac{I_{on}}{I_{off}}$ ratio of logic transistors, it also has been well known that a pronounced shift in the threshold voltage ($\Delta V_{th}$) can be achieved by incorporating the FE into the gate stack \cite{doi:10.1063/1.351910}. In this case the current-voltage characteristics exhibits a hysteresis. Such an unique property makes FeFETs a promising candidate for memory applications due to its non-destructive read, fast read and write operation, and non-volatility \cite{doi:10.1063/1.351910,1477770}. As a result, it is of importance to have a clear picture for the underlying physics behind FeFETs for both memory and logic operation regimes. Unlike previous works focusing solely on the particular application: (i.e., presuming that initially FeFETs are in the NC regions \cite{doi:10.1021/nl071804g,5783903,7588064,7590009} or a clear hysteresis loops are established in FE oxides \cite{1176521}), this paper does not assume any particular operation region initially and provides a unified picture of how FeFET operation transitions from memory to logic devices due to the change of thermodynamic free energy profiles. This picture is verified by numerical simulations incorporating MOS electrostatics 
	and polarization dynamics self-consistently.
	
	The rest of this paper is organized as follows. In Sec. II, capacitance enhancement using FE NC effects in series capacitors
	is demonstrated. Also the importance of the double-well free energy profile to memory devices is illustrated. Next, in Sec. III, a theoretical model describing both charge and current-voltage characteristics in FeFETs is presented in detail. Section IV shows the numerical results to support the key features observed from the free energy profiles and points out some issues that need to be addressed in FeFETs for both logic and memory applications in the future.  Conclusions are formulated in Section V.
	
	\section{Thermodynamic free energy in capacitors}
	In this section, charge distribution in several capacitors-in-series systems is determined from thermodynamic free energy point of view. We start from two DE capacitors in series to show the results from free energy aspect are consistent with those from the conventional circuit theory. Next a concept of super capacitor implemented by FE and DE capacitors in series is introduced using their free energy profiles. Finally, similarly to super capacitors, a superior MOS capacitor can be achieved by forming a FE thin film on top of a conventional MOS capacitor. As the FE thickness varies, a capacitor may or may not have the hysteresis effects. These two regimes are important in memory and logic applications.
	
	\begin{figure*}
		\begin{center}
		\subfloat[]{%
			\includegraphics[width=.12\linewidth]{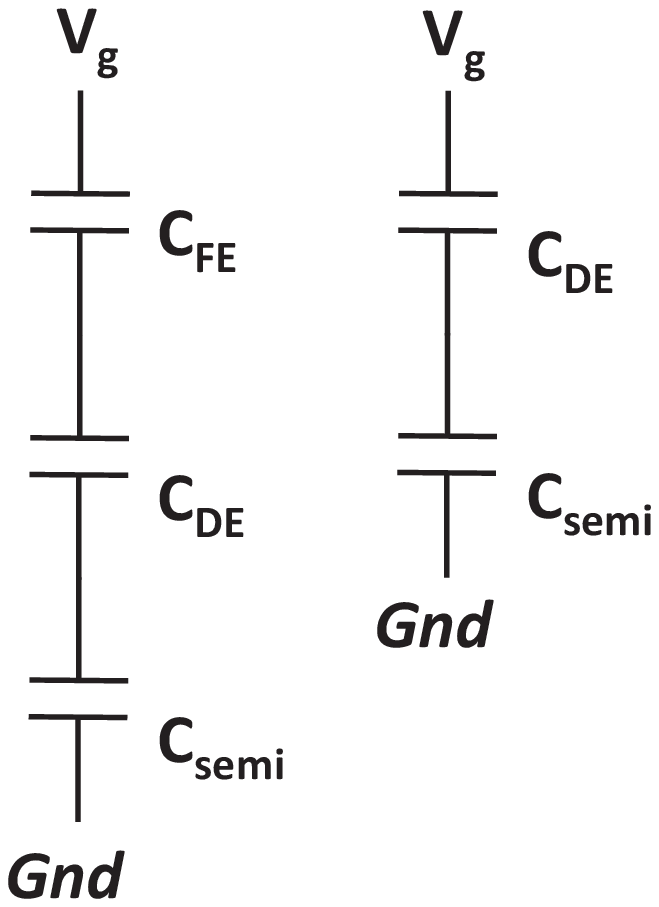}%
		}
		\subfloat[]{%
			\includegraphics[width=.35\linewidth]{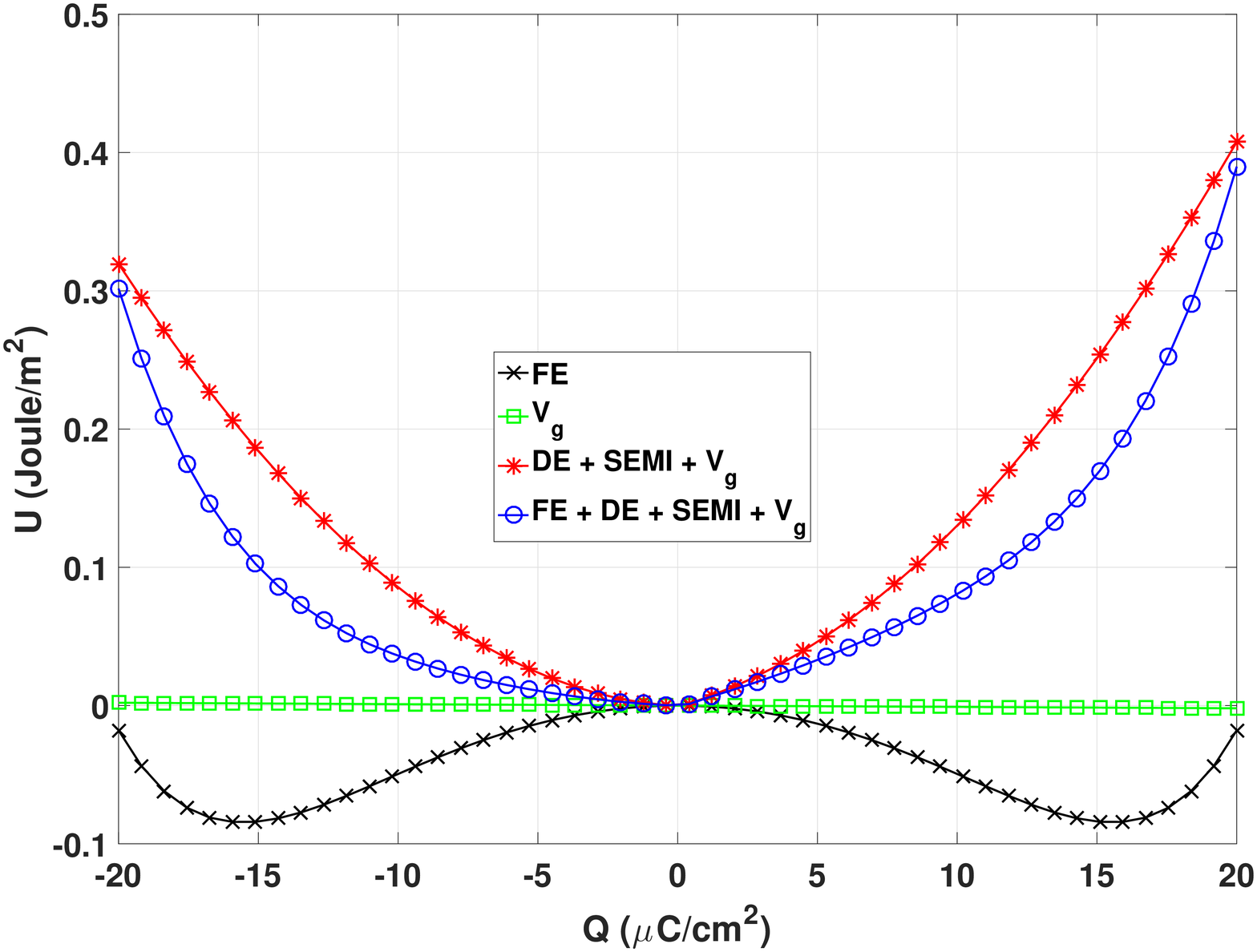}%
		}\\
		\subfloat[]{%
			\includegraphics[width=.35\linewidth]{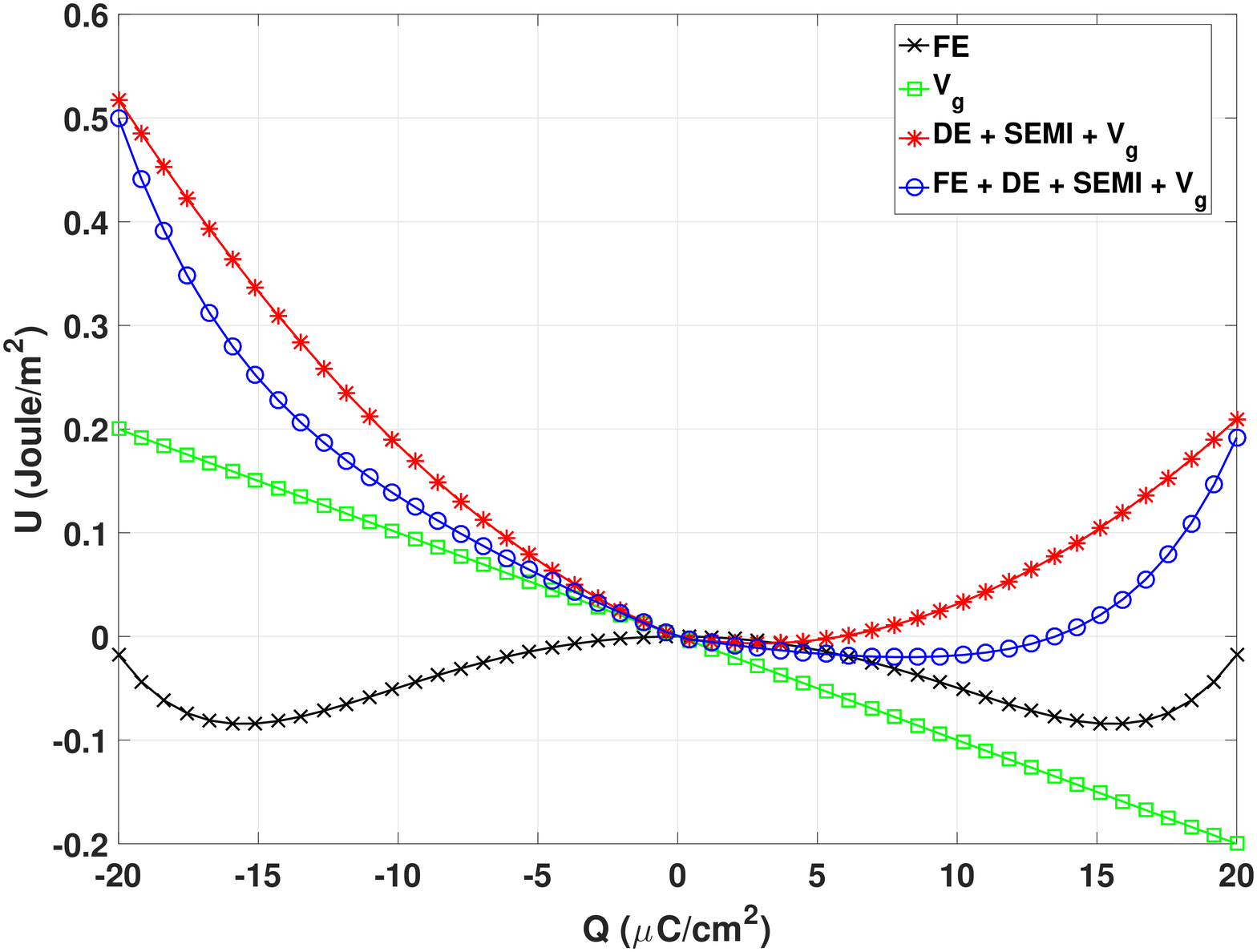}%
		}
		\subfloat[]{%
			\includegraphics[width=.35\linewidth]{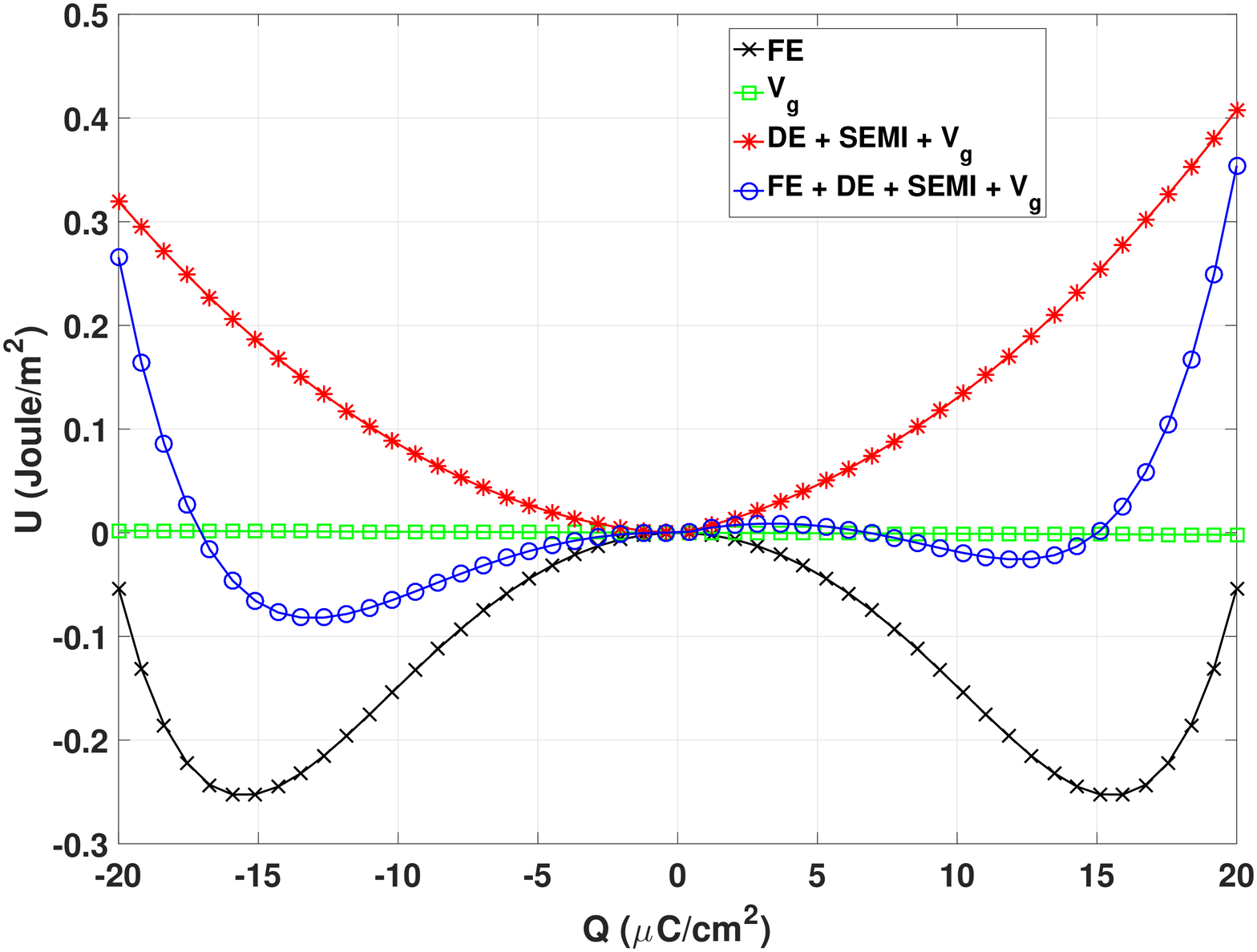}%
		}
		\caption{(a) Schematics of a series of FE, DE, and semiconductor capacitors as well as a series of DE and semiconductor capacitors. Free energy profiles of a traditional MOS stack (red curve) and a FeMOS stack (blue curve) with (b) $t_{FE}=5$nm, $t_{DE}=0.5$nm, $V_{g}=10$mV, (c) $t_{FE}=5$nm, $t_{DE}=0.5$nm, $V_{g}=1$V, and (d) $t_{FE}=15$nm, $t_{DE}=0.5$nm, $V_{g}=10$mV. The green and black curves designate contributions to the free energy by the applied voltage and the FE capacitor, respectively. The doping density of semiconductor is $10^{18}$ cm$^{-3}$.}
		\label{fig2}
		\end{center}
	\end{figure*}
	
	\subsection{Two dielectric capacitors in series}
	When two DE capacitors are connected in series, the total free energy of the system, $U_{tot}$, can be written as
	
	\begin{eqnarray}
	U_{tot} = U_{1} + U_{2} + U_{V},
	\label{eq1}
	\end{eqnarray}
	where $U_{1}=\frac{Q^{2}}{2C_{1}}$ and $U_{2}=\frac{Q^{2}}{2C_{2}}$ are the energies stored in each capacitor, respectively, and $U_{V}=-QV$ is the energy due to the applied voltage, $V$. $Q$ is the free charge on metallic plates, and $C_{1}$ and $C_{2}$ are capacitance of the two capacitors. In equilibrium ($V=0$), if we express $U_{tot}$ in terms of free energy and total capacitance, $C_{tot}$, Eq. \ref{eq1} becomes
	
	\begin{eqnarray}
	\frac{Q^{2}}{2C_{tot}}=\frac{Q^{2}}{2C_{1}} + \frac{Q^{2}}{2C_{2}}.
	\label{eq2}
	\end{eqnarray}
	From Eq. \ref{eq2}, it can be seen that the inverse of total capacitance for two capacitors in series is simply equal to the sum over the inverse of that for each capacitor. That is, $\frac{1}{C_{tot}}=\frac{1}{C_{1}}+\frac{1}{C_{2}}$, which is consistent with the well-known circuit theory result. With voltage applied, the total energy becomes
	\begin{eqnarray}
	U_{tot} = \frac{Q^{2}}{2C_{tot}} - QV,
	\label{eq3}
	\end{eqnarray}
	In Eq. \ref{eq3}, the steady-state charge $Q$ at a given voltage can be found via the extreme condition $\frac{\partial U_{tot}}{\partial Q}=0$. This resuts in $Q=C_{tot}V$ which is also consistent with the circuit theory. Consequently, the total energy for series capacitors is simply the sum over the free energy of individual components.
	
	\subsection{Ferroelectric and dielectric capacitors in series}
	As a capacitor is fabricated by depositing a FE thin film on top of a DE layer, its total energy is similar to the previous case
	
	\begin{eqnarray}
	U_{tot} = U_{FE} + U_{DE} + U_{V},
	\label{eq4}
	\end{eqnarray}
	where 
	\begin{eqnarray}
	U_{FE}=t_{FE}\left(\alpha_{1}Q^{2}+\alpha_{11}Q^{4}+\alpha_{111}Q^{6}\right)
	\end{eqnarray} 
	is the free energy of a FE layer under the single-domain approximation \cite{PhysRevB.20.1065}, $U_{DE}=\frac{Q^{2}}{2C_{DE}}$ is the free energy of a DE layer, and $U_{V}=-QV$ is the free enegry due to the applied voltage. Note that $U_{FE}$ here is expressed assuming that $Q=\epsilon_{0}E_{FE}+P\simeq P$ with $\epsilon_{0}$ being the vacuum dielectric constant, $E_{FE}$ being the electric field across a FE thin film, and $P$ being the FE polarization. In the following results for free energies, lead zirconate titanate (PZT) and SiO$_{2}$ are used as example FE and DE materials to illustrate the conceptual idea, and the corresponding parameters are given in Table. \ref{tab1}. In equilibrium ($V=0$), the free energy profiles for each component are shown in Fig. \ref{fig1}(b), where no charge is accumulated on either DE or DE-FE capacitors. That is, the minima of free energy are located at zero charge. If we focus on the curvatures near the minima of free energy profiles, it can be seen that the DE-FE capacitor has larger capacitance compared to the regular DE as obvious from the definition $C=\left(\frac{\partial^{2} U}{\partial Q^{2}}\right)^{-1}$. When the same voltage is applied to both capacitors, as shown in Fig. \ref{fig1}(c), it can be seen that larger charge is induced in a DE-FE capacitor. This can be alternatively understood either as a smaller curvature near the free energy minimum of a DE-FE capacitor or as the global free energy minimum located at a larger value of charge. As a result, a better capacitor (or super capacitor) can be achieved by simply adding a FE layer on top of a regular DE capacitor or having a FE capacitor with inevitable dead layers as discussed by several previous works \cite{PhysRevB.63.132103,doi:10.1063/1.2408650,PhysRevLett.99.227601,Stengel2009}. Note that the idea of a super capacitor works only when the total free energy profile has only one global minimum. That is, the dominant component in the overall system is still the DE and not the FE layer. Therefore as the FE thickness is increased, the total free energy profile get a more double-well-like shape as shown in Fig. \ref{fig1}(d), hysteresis effects start showing up in DE-FE capacitors.
	
	\subsection{Ferroelectric and metal-oxide-semiconductor capacitors in series}
	Similar to the concept of a super capacitor mentioned above, it is also possible to achieve a superior MOS capacitor by having a FE thin film on top of a regular MOS capacitor (FeMOS). As can be seen in Fig. \ref{fig2}(b), the free energy associated with a nonlinear semiconductor capacitor ($U_{semi}=\frac{Q\psi_{s}}{2}$ with $\psi_{s}$ being the surface potential drop within a semiconductor) is now included in the total free energy profile. Here PZT, SiO$_{2}$, and Si are used as FE, DE, and semiconductor materials to illustrate the concept. The parameters for SiO$_{2}$ and Si can be found in Table. \ref{tab1}.  The total free energy of the system is given as
	
	\begin{eqnarray}
	U_{tot} = U_{FE} + U_{DE} + U_{semi} + U_{V_{g}},
	\label{eq5}
	\end{eqnarray}
	Near equilibrium ($V_{g}\simeq0$), with a proper FE thickness, it can be seen in Fig. \ref{fig2}(b) that the capacitance of a FeMOS capacitor is greater than that of the regular MOS one. This improvement is mainly due to the global minimum of a FeMOS energy located where the curvature of a FE thin film energy is negative. As a result, at a given gate voltage, more charge can be accumulated on a FeMOS capacitor compared to other capacitors as shown in Fig. \ref{fig2}(c). Hence, a FET structure with a FeMOS capacitor can generate more drain currents if no significant mobility degradation is produced by the FE thin film. This benefits both high-performance and low-power logic applications since it is possible to control the charge boost at different gate-voltage regions. As the FE material becomes thicker, the free energy profile of a FeMOS capacitor starts transforming from one global minimum to double local minima (see Fig. \ref{fig2}(d)), and thus the hysteresis effects become more pronounced. In general, a good memory device can be built based on significant hysteresis effects in FeMOS capacitors \cite{7827698}.
	
	\section{Theoretical Model}
\begin{figure}
		\begin{center} 
		\includegraphics[width=0.6\linewidth]{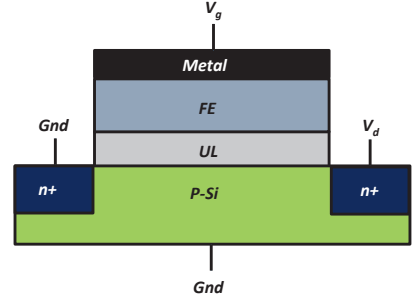}
		\caption{Schematic of an n-FeFET structure described by the theoretical model introduced in Sec. III. Along the gate direction, the device is composed of a metal, a FE layer, a DE layer, and a p-type semiconductor. Highly-doped n-type semiconductors are used for the source and drain.}
		\label{fig3}
		\end{center} 
	\end{figure}

	In this section, we introduce a mathematical model for FeFETs combining 1-D MOS electrostatics (along the gate direction) with FE polarization dynamics to justify the idea we discussed in the previous section. Here we use n-FETs as an example for a proof of concept, though a qualitatively same behavior occurs in p-FETs as well. The FeFET structure simulated below is shown in Fig. \ref{fig3}, where a gate stack is composed of metallic, FE, DE, and p-type semiconducting materials and highly n-doped semiconductors are used for the source and drain. The material parameters are also summarized in Table. \ref{tab1}.
	
\begin{figure}
		\begin{center} 
		\includegraphics[width=0.9\linewidth]{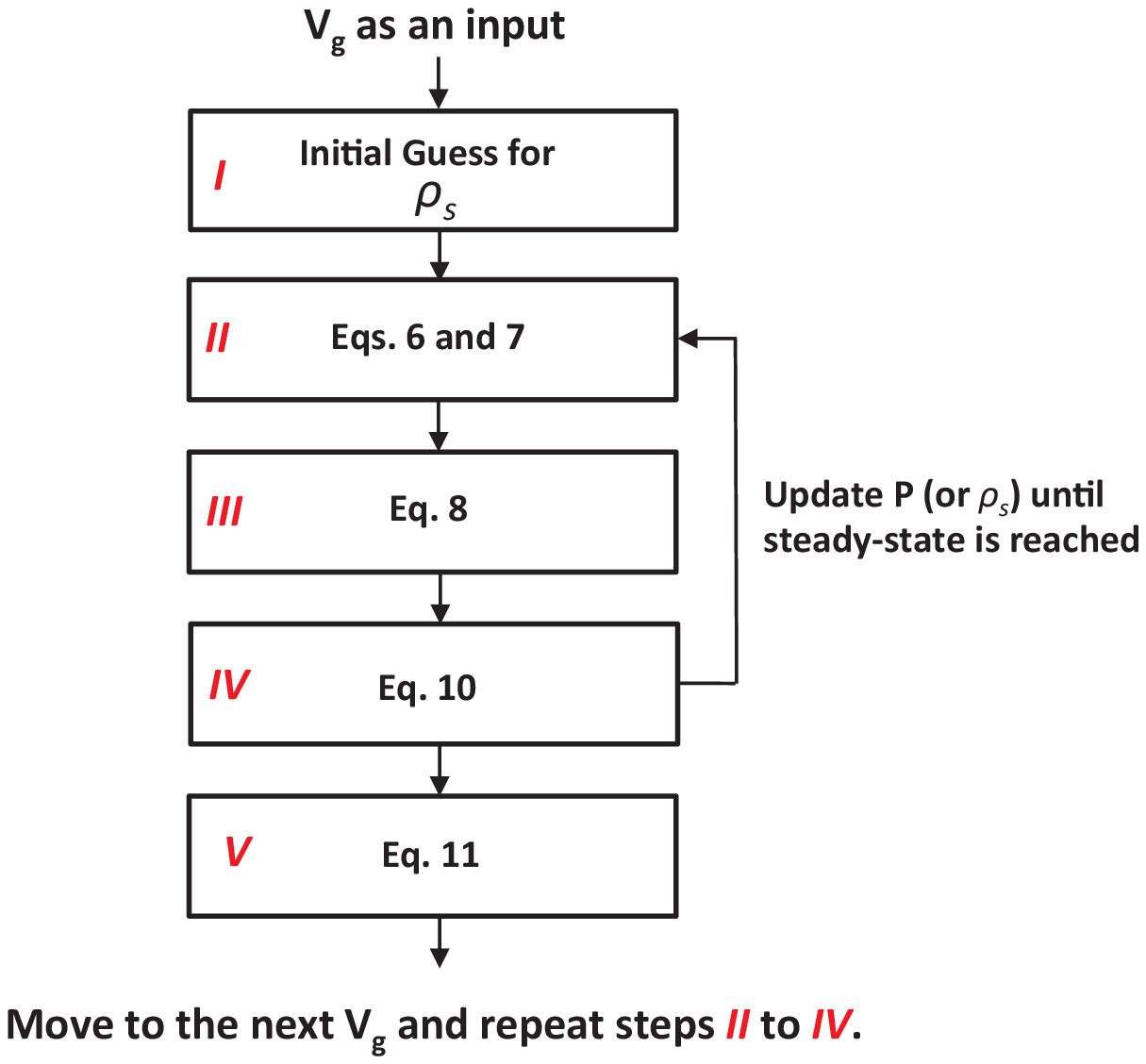}
		\caption{Numerical procedure for steady-state simulations in FeFETs}
		\label{fig4}
		\end{center} 
	\end{figure}

	To simulate a FeFET under steady-state conditions which correspond to the local or global minima in free energy profiles, at a given voltage, we start from an initial guess for $\rho_{s}$, the free charge density on the metal side. By assuming that the electrical displacement along the gate direction is continuous, the following equations are satisfied.
	
	\begin{eqnarray}
	\rho_{s}=P=\epsilon_{DE}\epsilon_{0}E_{DE}=-Q_{s},
	\label{eq6}
	\end{eqnarray}
	where $\epsilon_{DE}$ is the relative dielectric constant of a DE layer, $E_{DE}$ is the electric field across the DE layer, and $Q_{s}$ is the charge in the semiconductor channel. Note that $Q_{s}$ can be related to the semiconductor surface potential, $\psi_{s}$, through the following equation \cite{Taur:1998:FMV:291188}
	
	\begin{eqnarray}
	Q_{s} &=& \pm\sqrt{2\epsilon_{semi}\epsilon_{0}k_{B}TN_{a}}\left[\left(\frac{n_{i}}{N_{a}}\right)^{2}\left(e^{\frac{e\psi_{s}}{k_{B}T}}-\frac{e\psi_{s}}{k_{B}T}\right.\right. \nonumber \\ 
	&-& \left.1\right) + \left. \left(e^{-\frac{e\psi_{s}}{k_{B}T}}+\frac{e\psi_{s}}{k_{B}T}-1\right)\right]^{\frac{1}{2}},
	\label{eq7}
	\end{eqnarray}
	where $\epsilon_{semi}$ is the relative dielectric constant of a semiconductor, $k_{B}$ is the Boltzmann constant, $T$ is the temperature, $N_{a}$ is the p-type dopant concentration, $n_{i}$ is the intrinsic concentration of a semiconductor, and $e$ is the elementary charge. The $\pm$ sign in Eq. \ref{eq7} represents depletion/inversion or accumulation charge. Since the gate voltage, $V_{G}$, has to be shared within a FeMOS capacitor, the electric field across the FE layer, $E_{FE}$, can be obtained by Eq. \ref{eq8}.
	
	\begin{eqnarray}
	E_{FE}t_{FE} &=& eV_{G} - \left(\phi_{1} - \phi_{2} -E_{F} - \delta \right) - E_{DE}t_{FE} \nonumber \\
	&-& e\psi_{s} -\frac{\rho_{s}\lambda_{m}}{\epsilon_{m}\epsilon_{0}},
	\label{eq8}
	\end{eqnarray}
	where $t_{FE}$ is the FE thickness, $\phi_{1}$ and $\phi_{2}$ are conduction band discontinuities at the metal-FE and DE-semiconductor interfaces, respectively, $E_{F}$ is the Fermi energy in the metal, $\delta=E_{c}-\mu$ is the energy difference between the conduction band ($E_{c}$) and chemical potential ($\mu$) in the semiconductor, $t_{DE}$ is the DE thickness, $\lambda_{m}$ is the screening length of metal, and $\epsilon_{m}$ is the dielectric constant of the metal. Note that $\delta$ is obtained by using the Joyce-Dixon approximation \cite{6818391} given as
	
	\begin{eqnarray}
	\delta &=& k_{B}T\left[\ln\left(\frac{N_{c}}{n}\right) - \left(\frac{n}{\sqrt{8}N_{c}}\right)\right. \nonumber \\
	&+& \left.\left(\frac{3}{16}-\frac{\sqrt{3}}{9}\right)\left(\frac{n}{N_{c}}\right)^{2} \right],
	\label{eq9}
	\end{eqnarray}
	where $N_{c}$ is the effective density of states (DOS) in the conduction band given as $N_{c}=2\left(\frac{2\pi m^{*}_{e}k_{B}T}{h^{2}}\right)$ with $m^{*}_{e}$ being electron DOS effective mass and $h$ being Planck constant, and $n$ being the electron density in a bulk semiconductor. For simplicity, the Boltzmann approximation is assumed for $n\simeq\frac{n_{i}^{2}}{N_{A}}$. In Eq. \ref{eq8}, the potential energy drop within the metal is described under the Thomas-Fermi approximation \cite{PhysRevApplied.7.024005}. For a given $E_{FE}$. it is assumed that the FE polarization follows the Landau-Khalatnikov (LK) equation under the single-domain approximation given as \cite{PhysRevApplied.4.044014,7373582,PhysRevB.88.024106,PhysRevB.68.094113,PhysRevB.20.1065}
	
	\begin{eqnarray}
	\gamma\frac{\partial P}{\partial t}&=&-\frac{\partial}{\partial P}\left(\alpha_{1}{P^{2}}+\alpha_{11}P^{4}+\alpha_{111}P^{6}\right. \nonumber \\
	& & \left.-E_{FE}P\right),
	\label{eq10}
	\end{eqnarray}
	with $\gamma$ being the viscosity coefficient describing how fast the polarization in a FE thin film can follow an external electric field. As discussed in the previous section, the concept of superior MOS capacitors is established based on the state located right at the minimum of free energy profile, which corresponds to the steady-state solutions to the LK equation; that is, $\frac{\partial P}{\partial t}=0$. Hence, the iteration between Eqs. \ref{eq6} to \ref{eq10} is required to obtain the steady-state quantities such as charge and voltage drop in FeFETs for a given gate voltage, and the charge in a semiconductor channel (or $\psi_{s}$) can be converted into drain currents at a given drain-to-source voltage by using Pao and Sah’s integral given as \cite{Taur:1998:FMV:291188}
	
	\begin{eqnarray}
	I_{D} = \frac{e\mu_{n} W}{L}\int_{0}^{V_{DS}}\int_{\psi_{B}}^{\psi_{s}}\left(\frac{n_{i}^{2}}{N_{a}}\right)\frac{e^{-\frac{e\left(\psi-V\right)}{k_{B}T}}\partial\psi\partial V}{E\left(\psi,V\right)},
	\label{eq11}
	\end{eqnarray}
	with
	
	\begin{eqnarray} E\left(\psi,V\right)&=&\sqrt{\frac{2N_{a}k_{B}T}{\epsilon_{si}\epsilon_{0}}}\left\{\left(\frac{n_{i}}{N_{a}}\right)^{2}\left[ e^{\frac{-eV}{k_{B}T}}\left(e^{\frac{e\psi}{k_{B}T}}-1\right)\right.\right. \nonumber \\
	& &\left.\left. -\frac{e\psi}{k_{B}T}\right]+\left(e^{\frac{-e\psi}{k_{B}T}}+\frac{e\psi}{k_{B}T}-1\right)\right\}^{\frac{1}{2}}
	\label{eq12}
	\end{eqnarray}
	being the channel electric field, $V_{DS}$ being the voltage across the source and drain terminals, $\mu_{n}$ is the effective electron mobility, $W$ and $L$ being the channel width and length, respectively, and $\psi_{B}$ being an infinitesimal number. Figure \ref{fig4} summarizes the procedure (flow-chart) to simulate FeFETs. Note that for the simulation results shown in the next section, we also include MOSFETs with high-k gate DE as a baseline (see the parameters in Table. \ref{tab1}), and the corresponding changes in the model are simply replacing $P$ in Eq. \ref{eq6} with $\epsilon_{high-k}\epsilon_{0}E_{high-k}$ and that no iteration is required.
	
	\begin{table}
		\caption{\label{tab1} Material parameters for thermodynamic free energy profiles and self-consistent calculations between 1-D MOS electrostatics and polarization dynamics shown in this work.}
		\begin{tabular}{c c c}
			\hline
			\hline
			Symbol & Quantity & Value \\
			\hline
			$\alpha_{1}$ &  & $-1.05\times10^{9}\frac{m}{F}$ \cite{7458805} \\
			$\alpha_{11}$ & Landau coefficient & $10^{7} \frac{m^{5}}{C^{2}F}$ \cite{7458805} \\
			$\alpha_{111}$ & & $6\times10^{11}\frac{m^{9}}{C^{4}F}$ \cite{7458805} \\
			$\epsilon_{SiO_{2}}$ & SiO$_{2}$ relative dielectric constant & $3.9$ \cite{El-Kareh:2009:SDP:1538309} \\
			$\epsilon_{HfO_{2}}$ & HfO$_{2}$ relative dielectric constant  & $22$ \cite{7496855} \\
			$\epsilon_{Si}$ & Si relative dielectric constant & $11.7$ \cite{El-Kareh:2009:SDP:1538309} \\
			$n_{i}$ & Intrinsic concentration & $1.5\times10^{10}\frac{\sharp}{cm^{3}}$ \cite{Taur:1998:FMV:291188} \\
			$m^{*}_{e}$ & Electron DOS effective mass & $1.08m_{0}$ \cite{Taur:1998:FMV:291188} \\
			$\mu_{n}$ & Effective electron mobility & $200\frac{cm^{2}}{V\cdot sec}$ \cite{5783903} \\
			$W$ & Channel width & $100nm$ \\
			$L$ & Channel length & $20nm$ \\
			$\phi_{2}$ & Conduction band discontinuity & $3.2eV$ \cite{doi:10.1063/1.120473} \\
			& at DE-semiconductor interface &  \\
			$E_{f}$ & Fermi energy of metal & $4.5eV$ \cite{PhysRevLett.94.246802,doi:10.1063/1.3195075} \\
			$\lambda_{m}$ & Metal screening length & $0.5 \times10^{-10}m$\cite{PhysRevApplied.7.024005} \\
			$\epsilon_{m}$ & Metal relative dielectric constant & $2$ \cite{PhysRevApplied.7.024005} \\
			\hline
			\hline	
		\end{tabular}
	\end{table}
	
	\section{Results and Discussion}
	\subsection{Steady-state behavior of ferroelectric field-effect-transistors}
	In this section, first, the theoretical model introduced in the previous section is applied to simulate FeFETs and to justify the concept discussed from the free energy point of view. Without losing the essential physics in FeFETs, in the following simulations, $\psi_{1}$ is adjusted such that $\psi_{1}-\psi_{2}-E_{F}-\delta$ makes the off-current ($\sim 10^{-10}$A) for both Fe and high-k FETs occur at about the same gate voltage. Next, the steady-state and transient negative capacitance effects for FeFETs are discussed to provide a better understanding on device operations.
	
		\begin{figure*}
		\begin{center}
			\subfloat[]{%
				\includegraphics[width=.35\linewidth]{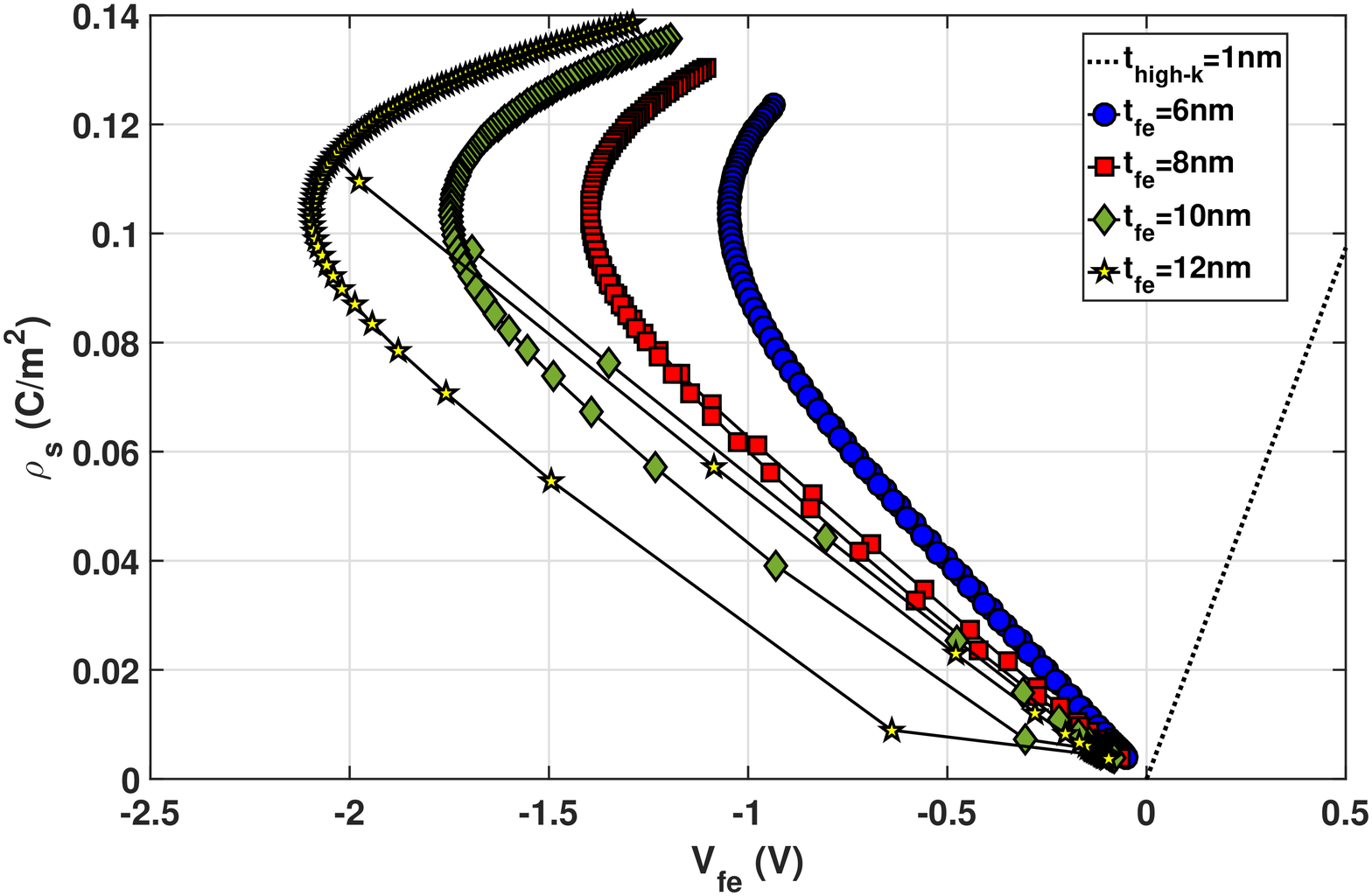}%
			}
			\subfloat[]{%
				\includegraphics[width=.35\linewidth]{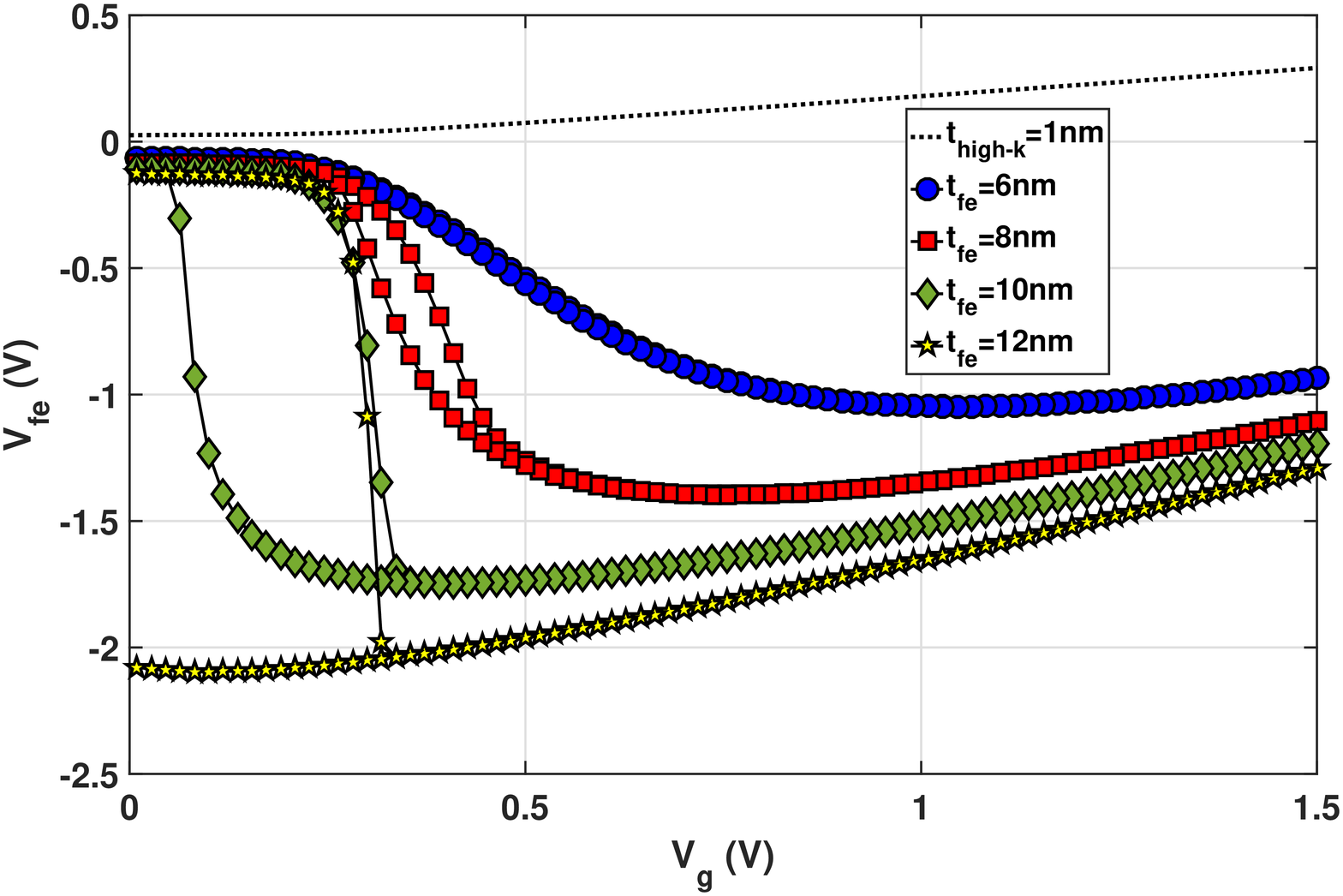}%
			}
			\subfloat[]{%
				\includegraphics[width=.35\linewidth]{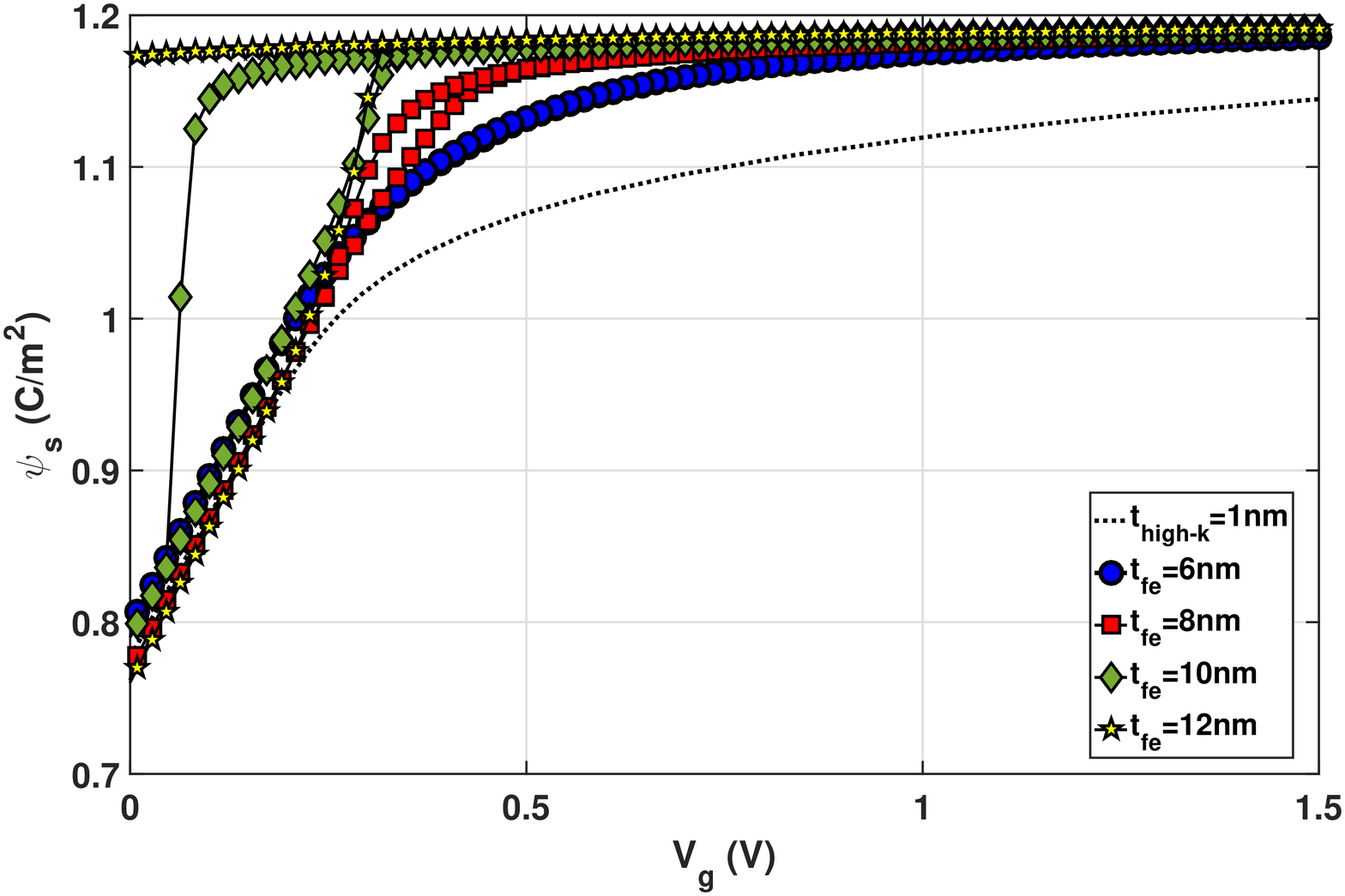}%
			}
			\caption{Comparison between FETs based on high-k DE and different FE thicknesses with the same under layer ($0.5$ nm): (a) charge as a function voltage across the FE, (b) voltage across the FE as a function of gate voltage, and (c) surface potential as a function of gate voltage. The high-k DE thickness is $1$nm. The viscosity coefficient, $\gamma$, is $5\times10^{2}$ m$\cdot$sec/F. The doping density of semiconductor is $10^{18}$ cm$^{-3}$.}
			\label{fig5}
		\end{center}
	\end{figure*}
	
	As discussed in Sec. II, the capacitance enhancement in a FeMOS capacitor mainly comes from the fact that the steady state of entire system is close to the negative capacitance region of the FE, which implies that the charge induced on the MOS capacitor is opposite to the voltage across the FE or the voltage across the FE is reduced as the gate voltage increases as shown in Figs. \ref{fig5}(a) and (b), respectively. As a result, for a given positive gate voltage, a larger surface potential drop in the semiconductor can be established in FeFETs compared to conventional high-k MOSFETs if the voltage across the FE layer is negative as shown in Fig. \ref{fig5}(c). Note that as the FE thickness is increased, the hysteresis effects in the devices become more pronounced because the overall free energy profiles are more double-well-like as explained previously.
	
\begin{figure*}
		\begin{center}
		\subfloat[]{%
			\includegraphics[width=.35\linewidth]{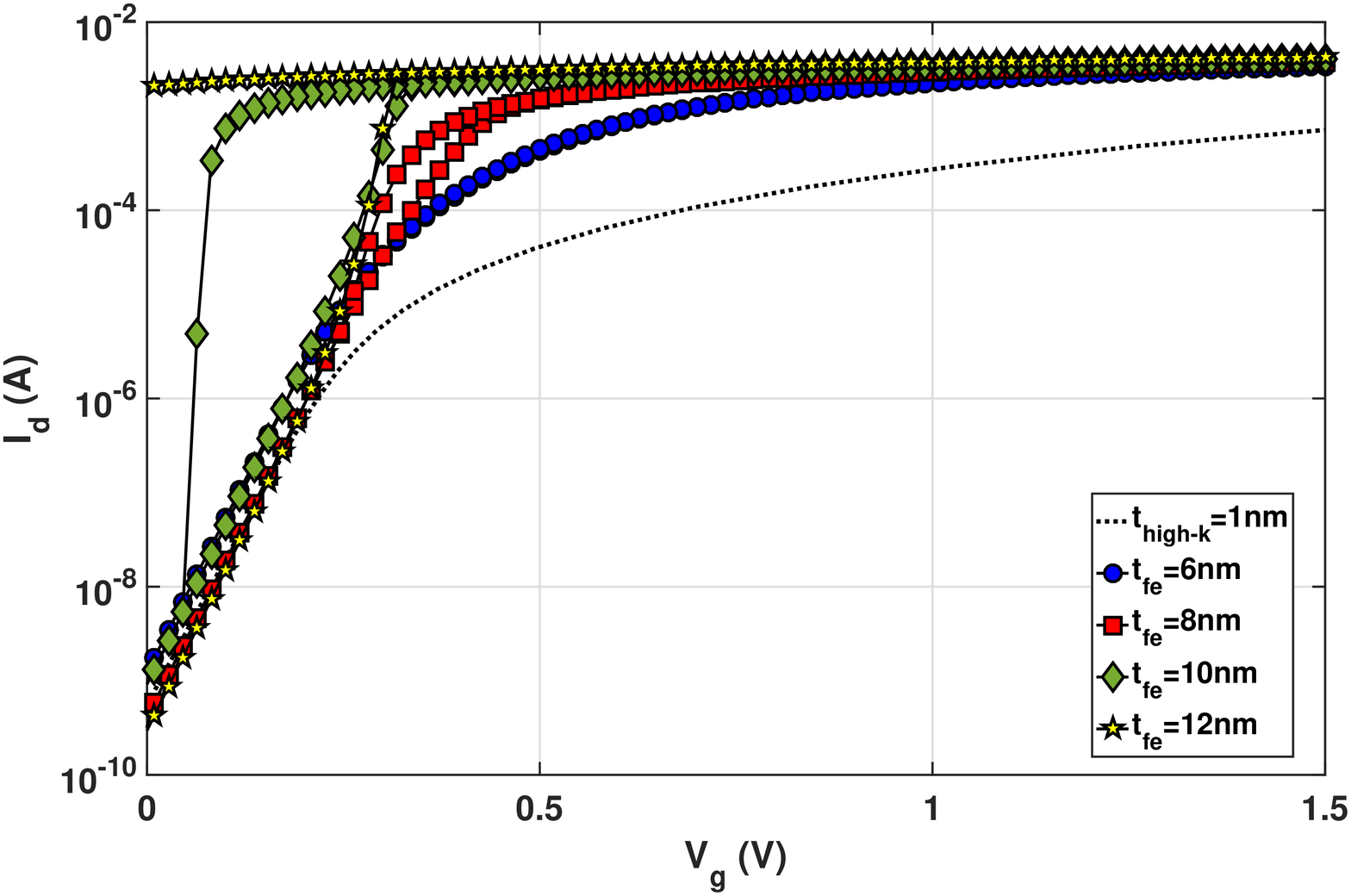}%
		} 
		\subfloat[]{%
			\includegraphics[width=.35\linewidth]{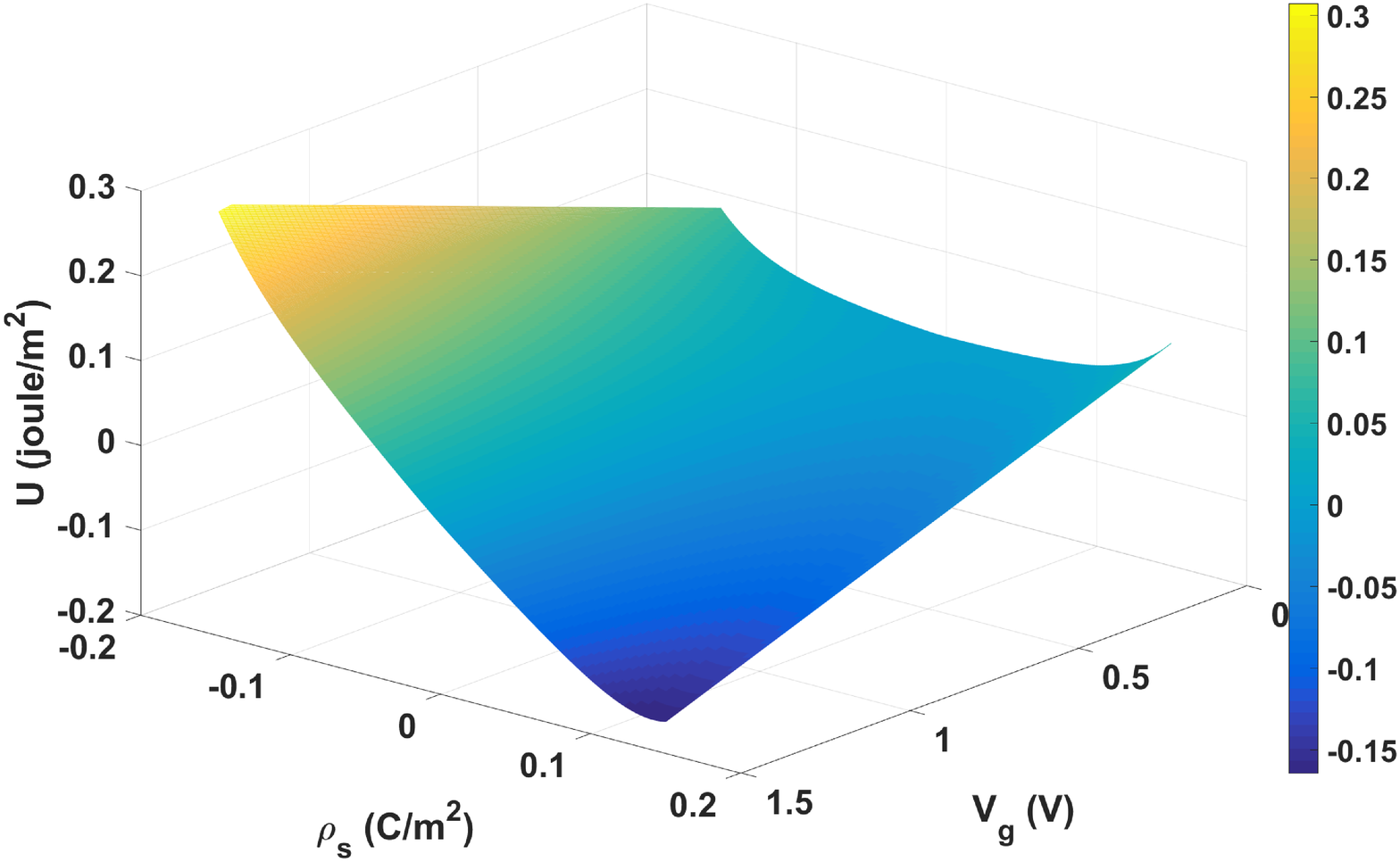}%
		}
		\subfloat[]{%
			\includegraphics[width=.35\linewidth]{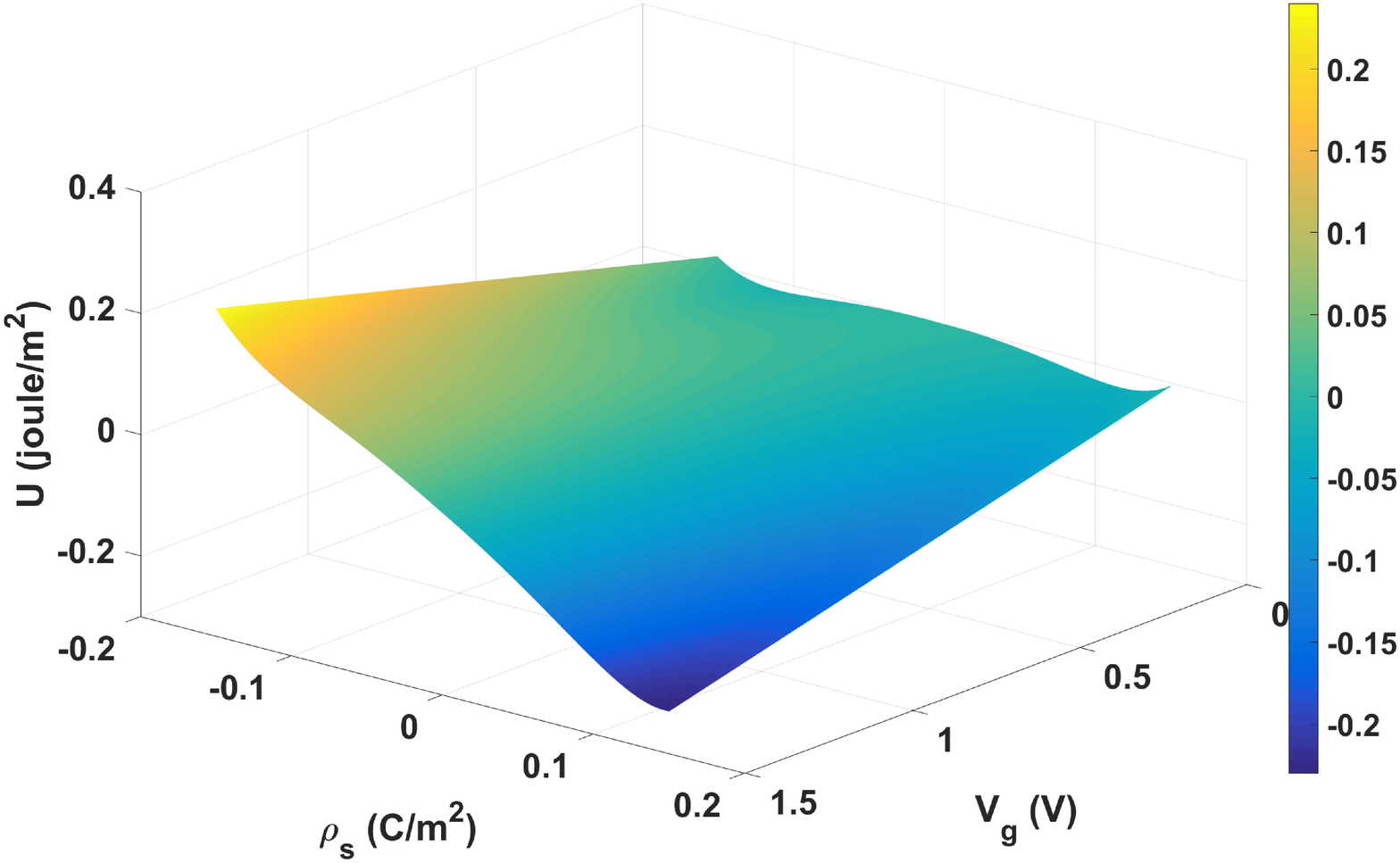}%
		}
		\caption{(a) Source-to-drain current vs. gate voltage in FeFETs with different FE thicknesses. The red curve with circles shows MOSFETs with 1nm high-k DE as a baseline. The under layer thickness is 0.5nm and the doping density of semiconductor is $10^{18}$ cm$^{-3}$ for all the cases. $V_{ds}=$50mV. (b) and (c) show free energy profiles as functions of both charge and gate voltage for FeFETs with thickness of $6$nm and $10$nm, respectively}
		\label{fig6}
		\end{center}
	\end{figure*}

	To convert the surface potential to drain currents, we assume electron mobility is not degraded significantly as the FE oxide is deposited on the DE layer; that is, the same mobility is used for all the cases in Fig. \ref{fig6}. Consequently, the current improvements in FeFETs come from the charge boost because of larger capacitance in the FeMOS stack. As shown in Fig. \ref{fig6}(a)  depending on the FE thickness, FeFETs can exhibit quite different I$_{d}$-V$_{g}$ characteristics and are potentially useful in both logic and memory applications. For thin FE films, a significant boost to on-currents compared to high-k MOSFETs is useful for high performance digital switches. As the FE film gets thicker, a great charge-boost in the subthreshold region appears at the expense of a weak hysteresis effect, which can decrease the gate voltage that is required to drive circuits. If we increase the FE thickness further, due to the double-well free energy profile as shown in Fig. \ref{fig2}(d), a large difference between threshold voltages during forward and backward gate voltage sweeps is created and can be useful in memory designs due to non-destructive readouts \cite{7827698}. Furthermore, free energy profiles as functions of both charge and gate voltage are shown in Figs. \ref{fig6}(b) and (c). We emphasize again the importance of a single free energy minimum over the entire gate voltage sweep to the hysteresis-free I$_{d}$-V$_{g}$ characteristics. In other words, sudden jumps in I$_{d}$-V$_{g}$ result from the fact that the negative curvature of free energy profile is not a thermodynamically stable state and thus the polarization switching is required. On the other hand, due to the depolarization field contributed from the DE layer, there is no negative curvature presented in the free energy profile, and therefore no abrupt switching is observed in I$_{d}$-V$_{g}$. Note that in real devices, charging defects at the FE/DE interface or within the DE layer due to a strong electric field may significantly reduce charge boost or memory window established by the FE layer \cite{7519093}. Consequently, having a proper ratio between FE and DE layers not only makes FeFETs work in the correct mode, but also minimizes the unwanted charging effects.  
	
	\subsection{Transient behavior in ferroelectric field-effect-transistors}
\begin{figure}
		\begin{center}
		\includegraphics[width=1\linewidth]{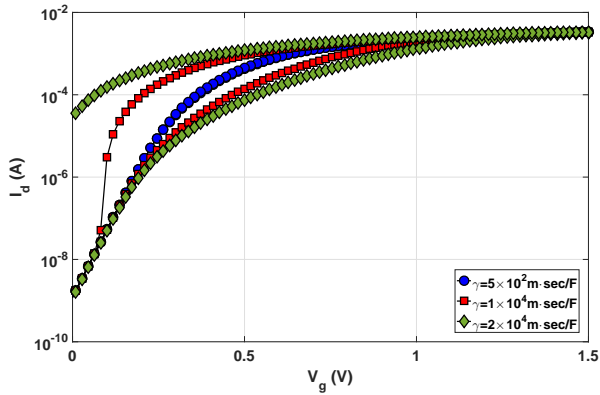}
		\caption{Source-to-drain current as a function of gate voltage in FeFETs ($t_{FE}= 6$nm) with different viscosity coefficients. The under layer thickness and doping density of semiconductor are $0.5$nm and $10^{18}$ cm$^{-3}$ for all the cases, respectively. $V_{ds}=$50mV.}
		\label{fig7}
		\end{center}
	\end{figure}

	As discussed previously, NC in FE thin films plays an extremely important role to enhance FeFET logic performance (i.e., charge-boost with no hysteresis). However, this improvement can only be achieved when the device reaches its steady-state condition under a given bias situation; that is, local or global minima located at the FE NC region in the free energy profile. How fast the steady state can be achieved is expressed by the viscosity coefficient ($\gamma$) in the LK equation, which largely depends on both material intrinsic and extrinsic properties such as FE domain wall nucleation and propagation rates as well as on thin film quality \cite{PhysRevB.68.094113}. As a result, the charge-boost in FeFETs may be significantly modified as the operation frequency goes too high. To capture this transient behavior correctly, a rigorous material calibration for $\gamma$ to FE switching dynamics is necessary. The effects of different viscosity coefficients on I$_{d}$-V$_{g}$ are given in Fig. \ref{fig7}, in which a larger viscosity coefficient implies a slower FE response. Note that for both forward and backward gate voltage sweeps, $1\mu$s is given for the polarization dynamics at each gate voltage in this work. As shown in Fig. \ref{fig7}, when the FE response becomes so slow such that the polarization cannot reach its steady state at a given voltage, the hysteresis effects become more significant even though a FE film is thin enough to form a single minimum in the free-energy profile.
	
	Next, we discuss the recent experimental demonstration on transient NC in FE capacitors \cite{Khan2015}. Note that the NC measured in such a resistance-capacitance (RC) setup is mainly from electrostatic effects, rather than thermodynamic energy profile. This is because the transient NC response shown in Ref. \cite{Khan2015} can be modulated significantly with different series resistance. In other words, the measured NC is simply due to the fact that the polarization switching in the FE is too fast such that the free charge provided from the external circuit cannot follow \cite{Catalan2015}. This phenomenon can be explained by the following electrostatic equation.
	
	\begin{eqnarray}
	V_{FE}=t_{FE}E_{FE}=t_{FE}\left(\frac{\rho_{ext}-P}{\epsilon_{0}}\right),
	\label{eq13}
	\end{eqnarray}
	Here $\rho_{ext}$ is the measured charge, whose frequency response is limited by external RC circuits. From Eq. \ref{eq13}, when $\rho_{ext}$ cannot follow to the polarization switching (i.e., $\rho_{ext}$ is assumed constant unlike $P$) the voltage across the FE can be reversed and thus transient NC is observed. Here we argue that this transient NC due to the mismatch between external charge and FE polarization responses cannot lead to an improvement in drain currents. This is mainly because charge in the semiconductor channel, which is proportional to the drain current, has to be always equal and opposite to that on the metal side due to charge neutrality. Note that the transient I$_{ds}$-V$_{gs}$ response discussed in Fig. \ref{fig7} results from the fact that FE polarization cannot reach its global minimum during a given gate voltage pulse, rather than the charge-polarization mismatch mentioned above, and thus a significant change in I-V characteristics can be induced.
	
	\section{Conclusion}
	We have presented a systematic understanding of FeFETs operating as memory and logic components from the thermodynamic point of view. 
	It is shown that, in order to have significant charge boost of channel charge and non-hysteresis behavior in FeFET-based high performance logic, a free energy profile with a single global minimum is required for the gate stack. On the other hand, for useful memory devices, significant $\Delta V_{t}$ in FeFETs can be established by double-well-like free energy profile. Furthermore the transition from a single global to a double local minima in free energy profiles can be achieved by varying the ratio between FE and DE thickness in FETs. These FeFET features deduced from thermodynamics are justified by numerical simulations including 1-D MOS electrostatics and FE polarization dynamics, and it is shown that, depending on the ratio between FE and DE thickness, the FeFET can potentially offer (i) higher drive current without hysteresis or steeper subthreshold swing with negligible hysteresis - for digital logic or (ii) significant memory windows - for memory applications. Also, the transient response of FeFETs and the effect on FET performance due to recent direct NC measurements are discussed.
	
	\bibliographystyle{IEEEtran}
	\bibliography{IEEEabrv,ieee_jxcdc}
	
\end{document}